\documentstyle[prl,aps]{revtex}
\begin{document}
\draft

\title{Resonant Photon-Assisted Tunneling Through a Double Quantum Dot:\\
An Electron Pump From Spatial Rabi Oscillations}
\author{C.~A.~Stafford$^{1,*}$ and Ned S.\ Wingreen$^2$}
\address{$\mbox{}^1$D\'{e}partement de Physique
Th\'{e}orique, Universit\'{e} de Gen\`{e}ve,
CH-1211 Gen\`{e}ve 4, Switzerland}
\address{$\mbox{}^2$NEC Research Institute, 4 Independence Way, Princeton,
New Jersey 08540}
\date{1 September 1995}
\maketitle

\makeatletter
\global\@specialpagefalse
\def\@oddhead{\underline{REV\TeX\mbox{ }3.0\hspace{11.7cm}
UGVA-DPT 1995 / 09-901}}
\let\@evenhead\@oddhead
\makeatother

\begin{abstract}
The time average of the fully  nonlinear
current through a double quantum dot, subject to an
arbitrary combination of ac and dc voltages,
is calculated exactly using the Keldysh
nonequilibrium Green function technique.
When driven on resonance, the system functions as an
efficient electron pump due to Rabi oscillation between the dots.
The pumping current is maximum when the coupling to the leads equals
the Rabi frequency.
\end{abstract}
\pacs{}

\tighten
\widetext

The spatial coherence of the electronic states in mesoscopic systems is
fundamental to understanding their dc transport properties \cite{dcreview}.
Recently, it has become possible to experimentally
investigate coherent effects in
time-dependent transport through mesoscopic systems \cite{pp}, opening
the possibility to study qualitatively new effects which depend in a crucial
way on the spatio-temporal coherence of the electronic states of a
time-dependent system.
While the phenomena of Ref.\ \onlinecite{pp} found
a natural explanation within linear response theory \cite{bpt,lhsd},
many time-dependent phenomena, such as electron pumps \cite{pump1},
photon-assisted tunneling \cite{tg,kou2,bs,sakaki,blick},
and lasers \cite{laser}, necessitate a nonlinear analysis.

In this Letter, we present a fully nonlinear treatment of
a novel electron pump based on a spatio-temporal
coherence effect: Rabi oscillation between states of a double quantum
dot. The double dot is modeled as two spatially separated
nondegenerate electronic orbitals,
each connected via a tunnel barrier to an electron reservoir (Fig.\ 1).
If the tunneling matrix element $\mbox{w}$
between the orbitals is small compared to their energy difference
$\Delta \epsilon=\epsilon_2-\epsilon_1$, the electrons
are highly localized on one orbital or the other,
inhibiting transport.  However, if the system is driven at a frequency
(or subharmonic)
corresponding to the energy difference $(\Delta \epsilon^2 +
4 \mbox{w}^2)^{1/2}$
between the time-independent eigenstates,
the electrons become
completely delocalized due to spatial Rabi oscillations.  If, as
shown in Fig.\ 1, the
reservoirs are biased in such a way that one reservoir can
donate electrons to the low-energy orbital ($\mu_L > \epsilon_1$)
and the other can accept electrons
from the high-energy orbital ($\mu_R < \epsilon_2$), the system
will then pump electrons from $\mu_L$ to $\mu_R$
with an efficiency that
can be made arbitrarily close to unity, {\it i.e.}, one electron transferred
for each photon absorbed.

We employ the Keldysh nonequilibrium Green function technique
\cite{wjm} to calculate the time-averaged current in response to an arbitrary
combination of ac and dc driving voltages, including finite coupling
to the leads.  The pumping current is found to be a maximum when the
coupling to the leads is equal to the Rabi frequency. Furthermore,
resonant features in the current are broadened by the coupling to the
leads, implying that additional levels of the dot contribute only
to a homogeneous background. Importantly, these results for the
double-dot system also apply to transport through a double
quantum well with negligible interface scattering.

The Hamiltonian of the double-dot system can be expressed as $H(t)=
H_{\mbox{dots}}(t) + H_1$, where
\begin{equation}
 H_{\mbox{dots}}(t) = \sum_{i=1}^2 \epsilon_{i}(t)
{\bf d}^{\dagger}_{i}
{\bf d}_{i}  + \mbox{w} ({\bf d}^{\dagger}_{2}
{\bf d}_{1} + \mbox{H.c.}),
\label{hoft}
\end{equation}
\begin{equation}
H_1 = \!\sum_{{\bf k},\ell \in L, R}\! \epsilon_{{\bf k}\ell}
{\bf c}^{\dagger}_{{\bf k}\ell} {\bf c}_{{\bf k}\ell} + \!\!\!\!\!\!
\!\sum_{{\bf k},\ell \in
\left\{\begin{array}{c} \!\!{\scriptstyle L, i=1} \\[-1mm]
\!\!{\scriptstyle R, i=2}
\end{array}\right.}
\!\!\!\!\!\!
\left(V_{{\bf k}\ell}
{\bf c}^{\dagger}_{{\bf k}\ell} {\bf d}_{i} + \mbox{H.c.}\right).
\label{h1}
\end{equation}
Here ${\bf d}^{\dagger}_{i}$ creates an electron in
the $i$th quantum dot and
${\bf c}^{\dagger}_{{\bf k}\ell}$ creates an electron
of momentum $\bf k$ in reservoir $\ell$.
For simplicity, spin is neglected and the external time
dependence is applied only to the dots \cite{sw},
$\epsilon_{i'}(t) = (-1)^{i'} (\Delta \epsilon + V \cos \omega t)/2$.
We do not explicitly treat interactions, but assume that
$\Delta \epsilon$ includes contributions both from electrostatic
confinement and Coulomb interaction within each dot.

Before discussing transport in a system connected to leads, it is useful
first
to consider the eigenstates of the closed system of two quantum dots
coupled capacitively to an ac voltage source, as described
by Eq.\ (\ref{hoft}).
The relevant eigenstates of a system such as (\ref{hoft}),
for which $H(t+2\pi/\omega)= H(t)$ is a periodic function of time,
are the eigenstates of the one-period evolution operator
$ U(t+2\pi/\omega,t) =  T
\{\exp [-\frac{i}{\hbar} \int_t^{t+2\pi/\omega} dt' \,H(t')]\}$.
For the double-dot system,
these states have the form \cite{ybz} \begin{equation}
\psi_{i}^{(j)} (t) = \exp (-iE_j t/\hbar) \varphi_{i}^{(j)}(t),
\label{qstate}
\end{equation}
where $E_j$ is the $j$th quasienergy, and
$\varphi_{i}^{(j)}(t+2\pi/\omega)=\varphi_{i}^{(j)}(t)$ is a Bloch function
whose components, $i=1,2$,  give the time-dependent
amplitudes on the two quantum dots.  The eigenvalue
problem defined by Eqs.\ (\ref{hoft}) and (\ref{qstate}) must in general
be solved numerically \cite{hh} because $[ H(t), H(t')] \ne 0$.
However, in the experimentally interesting case of strongly localized dc
eigenstates, $\mbox{w} \ll \Delta \epsilon$,
Eqs.\ (\ref{hoft}) and (\ref{qstate}) can be
solved analytically by expanding $U(t,t')$ to linear order in
$\mbox{w}$:  At the $N$-photon resonance,
$N\hbar\omega = \sqrt{\Delta \epsilon^2 + 4 \mbox{w}^2}
\simeq \Delta \epsilon$,  one finds for the quasienergies
\begin{equation}
E_{\pm} = \Delta \epsilon /2 \pm \mbox{w} \mbox{J}_N (V/\hbar \omega),
\label{qeloc}
\end{equation}
where $\mbox{J}_N$ is the Bessel function of order $N$.
For a small detuning $\delta \omega$ away from the $N$-photon resonance,
the occupancy of dot 1 in state $E_{\pm}$ is
$|\psi_{1}^{(\pm)}|^2 = [1 + (x \mp \sqrt{x^2+1})^2]^{-1}$,
where $x=\hbar\omega \sin(\pi N \delta \omega/\omega)/[2 \pi \mbox{w}
\mbox{J}_N (V/\hbar\omega)]$.  The quasienergy eigenstates are thus completely
delocalized on resonance ($|\psi_{1,2}^{(\pm)}|^2 =1/2$).

Qualitatively, the behavior near resonance for $\mbox{w} \ll \Delta
\epsilon$ can be understood
in terms of the hybridization of the
electronic orbital on one dot with the $N$th sideband of the electronic
orbital on the other dot (Fig.\ 2). For example, in the voltage frame
in which $\epsilon_2= \Delta \epsilon$ is
independent of time, the energy spectrum of the first dot in the absence
of tunneling has peaks at $E = N \hbar \omega$
with amplitudes
$\mbox{J}_N (V/\hbar \omega)$, as discussed in Refs.\
\cite{tg,wjm}.  When the energy of one of these sidebands
coincides with $\epsilon_2$, interdot tunneling will hybridize the two
orbitals into two delocalized combinations. The effective
coupling between orbitals is the product of $\mbox{w}$ and the sideband
amplitude, leading to the energy splitting in  (\ref{qeloc}).
An electron placed on one of the dots at resonance will therefore
oscillate back and forth between the dots at the Rabi
frequency $\Omega_R/\hbar = 2 ({\rm w}/\hbar) \mbox{J}_N(V/\hbar\omega)$.
It should be emphasized that although the quasienergy states are delocalized
on resonance, their energy spectrum remains
spatially asymmetric, centered near
$\epsilon_1 = 0$ on dot 1 and near $\epsilon_2 = \Delta \epsilon$
on dot 2; the delocalized states
must be thought of as coherent superpositions of states of the coupled
electron-photon system.

The coupling of the double-dot system to the reservoirs is characterized by
the parameters
$\Gamma^{L/R}(\epsilon) = 2\pi \sum_{{\bf k},\ell \in L/R}
|V_{{\bf k}\ell}|^2 \delta(\epsilon - \epsilon_{{\bf k}\ell})$.
In order to obtain
an analytic solution for the nonequilibrium time-dependent transport, we
consider the case where $\Gamma^L(\epsilon)
= \Gamma^R(\epsilon) = \Gamma$ is independent of
energy.  The expectation value of the current through the left barrier
can then be expressed using the formalism of Ref.\ \onlinecite{wjm} as
\begin{equation}
J_L (t)  =
\frac{-2e\Gamma}{\hbar} \int_{-\infty}^t dt' \int \frac{d\epsilon}{
2 \pi} \, \mbox{Im}
\left\{ e^{-i\epsilon(t'-t)}
\left[G_{11}^{<}(t,t') + f_L (\epsilon)
G_{11}^r(t,t')\right]\rule[-0.2cm]{0cm}{.4cm}\right\},
\label{joft}
\end{equation}
where $G_{i i'}^<(t,t') \equiv i \langle {\bf c}_{i'}^{\dagger}(t')
 {\bf c}_{i}(t) \rangle$
and $G_{i i'}^r(t,t') \equiv
-i\theta(t-t') \langle \{{\bf c}_{i'}^{\dagger}(t'),
 {\bf c}_{i}(t) \}\rangle$
are Green functions describing propagation within the double-dot
system in the presence of coupling to the leads.
The retarded Green function can be expressed simply in terms of the
quasienergy eigenstates as $G_{i i'}^r(t,t') = -i\theta(t-t')
\exp[-\Gamma (t-t')/2] \sum_j \psi_i^{(j)}(t) \psi_{i'}^{(j)\ast}(t')$.
Given $G^r$, the other Green function
$G^<$ can be determined via the Keldysh relation
\cite{wjm}, which allows the time-average of
$J_L(t)$ to be expressed in terms of the Fourier components of the quasienergy
Bloch functions as
\begin{eqnarray}
\bar{J} &=& \frac{e\Gamma}{\pi\hbar} \left[\int d\epsilon
f_L (\epsilon) \sum_{j,n}\mbox{Im}
\left\{ \frac{|\varphi_{1n}^{(j)}|^2}
{n \hbar\omega + E_{j} - \epsilon  - i\Gamma/2} \right\} \right.
\nonumber \\
& & \left. - \frac{\Gamma}{2} \sum_{\ell \in L,R} \int d\epsilon
f_{\ell}(\epsilon)\sum_{\stackrel{\scriptstyle i',j,j'}{n,n',m}}
\frac{\varphi_{i'n}^{(j)\ast}\, \varphi_{1n'}^{(j)} \,
\varphi_{1,n'+m}^{(j')\ast} \, \varphi_{i',n+m}^{(j')}}
{ (n \hbar\omega + E_j - \epsilon - i\Gamma/2)
[(n+m) \hbar\omega + E_{j'} - \epsilon + i\Gamma/2]} \right] ,
\label{jave}
\end{eqnarray}
where
\begin{equation}
\varphi_{i'n}^{(j)} = \frac{\omega}{2\pi} \int_{-\pi/\omega}^{\pi/\omega}
\!\! dt e^{inwt}\varphi_{i'}^{(j)}(t).
\label{phidef}
\end{equation}
Eq.\ (\ref{jave}) is an exact result for the time-averaged current,
valid for arbitrary gate and bias voltages.

Figure \ref{fig3} shows the time-averaged current for the case
$\mu_L=\mu_R=0$ for  several ac driving voltages,
calculated via Eq.\ (\ref{jave}) in the limit of zero temperature.
A series of peaks
in $\bar{J}$ are evident, occurring at the frequencies $\omega_N =
(\Delta \epsilon^2 + 4\mbox{w}^2)^{1/2}/N\hbar$,
corresponding to the delocalization transitions.
At the $N$th peak,
the dc current flows in response to resonant $N$-photon-assisted tunneling:
When the electron is on dot 1, it has an energy
$\simeq -\Delta\epsilon /2 < \mu_L$, and can not tunnel into reservoir $L$.
In performing a Rabi oscillation to dot 2, the electron absorbs
N photons, giving it an energy
$\simeq -\Delta\epsilon /2 + N\hbar\omega \simeq \Delta\epsilon /2 > \mu_R$;
the electron can thus tunnel from dot 2 to reservoir $R$.
Subsequently, another electron can
tunnel from reservoir $L$ onto dot 1 and the process is repeated,
leading to a dc current.
Each of the two delocalized quasienergy states contributes independently
to this current.  Consequently, one can resolve the
Rabi splitting between these states by sweeping one of the
chemical potentials.  For example,
the inset to Fig.\ \ref{fig3} shows sharp jumps in
$\bar{J}$ when $\mu_R$ crosses
the two quasienergies for the one-photon resonance
of the system, indicated as vertical dashed lines.
The spacing $\delta \mu_R$
between the two jumps is equal to the Rabi splitting
$\Omega_R \simeq 2\mbox{w}\mbox{J}_1(V/\hbar\omega)$ between the
quasienergy eigenstates.  The Rabi splitting can thus be resolved by
examining the $I$--$V$ characteristic of the electron pump.

In order to understand the heights and widths of the resonances in
$\bar{J}$, it is useful to consider the limit of strongly localized
orbitals $\mbox{w} \ll
\Delta\epsilon$ with weak driving $V \ll \hbar\omega$, so
that only resonant processes contribute to the current.
Using Eqs.\ (\ref{qeloc}) and (\ref{jave}) at zero temperature
and the fact that the
quasienergy eigenstates are completely delocalized on resonance, one obtains
the time-averaged current at the $N$-photon resonance,
\begin{eqnarray}
\bar{J}_{\mbox{res}}  = \frac{e\Gamma}{2h}
\left(\frac{\Omega_R^2}{\Omega_R^2 + \Gamma^2}\right)
& {\displaystyle \sum_{\sigma = \pm 1}} &
\left[\tan^{-1} \left(\frac{\mu_L - \epsilon_1 +
\sigma \Omega_R/2}{\Gamma/2}\right)
- \tan^{-1} \left(\frac{\mu_R - \epsilon_2 +
\sigma \Omega_R/2}{\Gamma/2}\right) \right.
\nonumber \\
& & \;\mbox{} + \left.\frac{\sigma \Gamma}{2\Omega_R} \ln \left(
\frac{(\mu_L - \epsilon_1 + \sigma \Omega_R/2)^2 + (\Gamma/2)^2}{
(\mu_R - \epsilon_2 + \sigma \Omega_R/2)^2 + (\Gamma/2)^2}\right)
\right].
\label{jave1}
\end{eqnarray}
Eq.\ (\ref{jave1}) shows explicitly that
each of the two delocalized quasienergy states contributes
independently to the resonant current.
For $\Gamma \ll \Omega_R$, the logarithmic term in Eq.\ (\ref{jave1})
is negligible, and the arctangents jump rapidly from $-\pi/2$ to $\pi/2$
when $\mu_{L,R}$ cross one of the quasienergies,
leading to the sharp jumps in $\bar{J}$ shown in the inset to Fig.\ \ref{fig3}.
Eq.\ (\ref{jave1})
predicts that $\bar{J}_{\mbox{res}}$ is not a monotonically
increasing function of the ac amplitude $V$, but reaches a maximum for
$V\sim \Delta \epsilon$ then decreases, due to the oscillatory character of the
Bessel function in $\Omega_R$.
This behavior is borne out in the exact solution.
The {\em inhibition} of transport at large ac amplitudes is one
feature which distinguishes true photon-assisted tunneling from
adiabatic electron transfer \cite{pump1}.

It is instructive to consider several limits
of Eq.\ (\ref{jave1}) with regard to the coupling to the leads $\Gamma$.
For $\mu_L - \epsilon_1$, $\epsilon_2 - \mu_R \gg \Gamma$, one finds
\begin{equation}
\bar{J}_{\mbox{res}} = \frac{e\Gamma}{2\hbar}\frac{\Omega_R^2}{
\Omega_R^2 + \Gamma^2}.
\label{jlim}
\end{equation}
For $\Gamma \ll \Omega_R$, $\bar{J}_{\mbox{res}}$
increases linearly with $\Gamma$, and
the resonances in $\bar{J}$ have an intrinsic width of
$\delta \omega_{\mbox{\scriptsize FWHM}} = 2 \Omega_R/N\hbar$.
$\bar{J}_{\mbox{res}}$ obtains a maximum
of $ e\Omega_R/4\hbar$ when the tunneling rate
to the leads is equal to the Rabi frequency.
In the limit $\Gamma\gg \Omega_R$, $\bar{J}_{\mbox{res}}\simeq
e\Omega_R^2/2\hbar\Gamma$ and the resonances are broadened
in energy by $\Gamma$ (and hence in frequency by
$\delta \omega_{\mbox{\scriptsize FWHM}}  = \Gamma/\hbar N$).
In this limit, the photon-assisted
tunneling is incoherent because the phase of the electron is randomized
on a time-scale short compared to the Rabi oscillations.
It is only in this limit, $\Gamma \gg \Omega_R$, that
$\bar{J}$ can be calculated via Fermi's golden rule using the lifetime
broadened density of states of the $N$th sideband,
as in the original calculation of Tien and Gordon \cite{tg}.
Eq.\ (\ref{jlim}) also implies that for $V < \hbar \omega$ the current at very
high-order resonances is exponentially suppressed compared to the current
at the 1-photon resonance, because
$\Omega_R \sim \mbox{J}_N(x) \sim (x/2)^N/N!$.
One can therefore generally neglect additional energy orbitals
within each quantum dot,
since even when in resonance the contribution to the current of an orbital
spaced by $\Delta E$ will be
exponentially small in $N = \Delta E/\hbar \omega$.

We find that the resonances in $\bar{J}$ are not broadened at finite
temperatures, provided $k_B T \ll \min|\mu_{\ell}-E_j|$.  A similar phenomenon
in dc resonant tunneling through a double quantum dot was recently observed
by van der Vaart {\it et al.} \cite{2dots}, underlining the analogy between
resonant photon-assisted tunneling between nondegenerate orbitals and dc
resonant tunneling through degenerate hybridized orbitals.
We find a reduction of $\bar{J}$ when
$k_B T\, \raisebox{-0.6mm}{$\stackrel{>}{\scriptstyle \sim}$}
\,\min|\mu_{\ell}-E_j|$,
indicating that a minimum energy $\sim k_B T$ must be dissipated in order for
the electron pump to operate at a maximal rate.

The limit $\Gamma \ll \Omega_R$ is of particular interest because it allows us
to estimate the intrinsic efficiency ${\cal E}$
of the electron pump, defined as the number of electrons
transferred per photon absorbed within the nanostructure, neglecting any
losses within the external ac voltage source.
For general $\Gamma_L,\Gamma_R \ll \Omega_R$,
Eq.\ (\ref{jlim}) becomes
$\bar{J}_{\mbox{res}} =e\Gamma_L\Gamma_R/\hbar(\Gamma_L+
\Gamma_R) \equiv J_{\mbox{max}}$.
This is the maximum dc current which can be
passed through a single pair of hybridized orbitals,
and signifies that {\em no} electrons traverse the system
in the opposite direction.  The intrinsic
efficiency of the electron pump at the one-photon resonance
is thus {\em unity} in this limit, since processes involving the
net absorption of $N\neq 1$ photons are negligible.
A lower bound on this efficiency
is ${\cal E} \geq \bar{J}/J_{\mbox{max}}$,
which becomes an equality as $\Gamma\rightarrow 0$.
One thus obtains $\lim_{\Gamma\rightarrow 0}
{\cal E} =1 - {\cal O}(\mbox{w}/\Delta\epsilon )^2$ at the one-photon
resonance, where the deviation from unity
stems from the fact that the dc eigenstates are not completely
localized.  For the
parameters of the inset to Fig.\ \ref{fig3}, one finds ${\cal E} \geq 0.93$,
even when pumping up potential gradients $\mu_R-\mu_L \sim \Delta\epsilon$.
This remarkably high efficiency stems from the coherent character
of photon-assisted tunneling in this system, in which resonant absorption
necessarily involves charge transfer.  Other electron pumps based on
photon-assisted tunneling in single dots\cite{kou2,bs} or intrawell optical
excitation \cite{sakaki}
do not share this feature.

Our results for the double-dot system can also be applied to vertical
transport through double quantum wells. If interface scattering
is negligible, each transverse mode is independent and can be
modeled by the same Hamiltonian used for the double-dot system.
A mode of transverse momentum ${\bf k}_{\bot}$
will contribute a current on resonance given by Eq.\ (\ref{jave1})
with $\epsilon_i \rightarrow \epsilon_i(0) + \hbar^2
{\bf k}_{\bot}^2/2m^{\ast}$,
where $m^{\ast}$ is the effective mass.
Integrating over transverse modes, including spin, one obtains, in the limit
$\mu_L - \epsilon_1(0)$, $\epsilon_2(0) - \mu_R \gg \Gamma$,
\begin{equation}
\bar{J}_{\mbox{2D}} =
\frac{e\Gamma}{2\hbar}\left(\frac{\Omega_R^2}{\Omega_R^2 + \Gamma^2}\right)
\frac{Am^{\ast}[\mu_L - \epsilon_1(0)]}{\pi\hbar^2},
\label{j2d}
\end{equation}
where $A$ is the area of the
2D electron gas.

In conclusion, we have obtained an exact solution for the time-average
of the fully nonlinear current driven through two quantum dots, each
coupled to an electron reservoir.
The system is found to function as an electron pump capable of
transporting electrons up large potential gradients with an efficiency
near unity due to resonant photon-assisted tunneling.
The pumping current is maximized when the coupling to the leads $\Gamma$
equals the Rabi frequency $\Omega_R$.  Since resonances
in the current are broadened by the coupling to the reservoirs,
the presence of additional energy orbitals contributes only to a
homogeneous background. These results also apply to transport through
double quantum wells with negligible interface scattering.

We thank Leo Kouwenhoven for raising our interest in this problem,
and Peter Wolff for valuable suggestions.
One of us (C.\ A.\ S.) acknowledges support from the Swiss National
Science Foundation.

\begin{figure}
\vbox to 8cm {\vss\hbox to 17cm
  {\hss\
    {\includegraphics{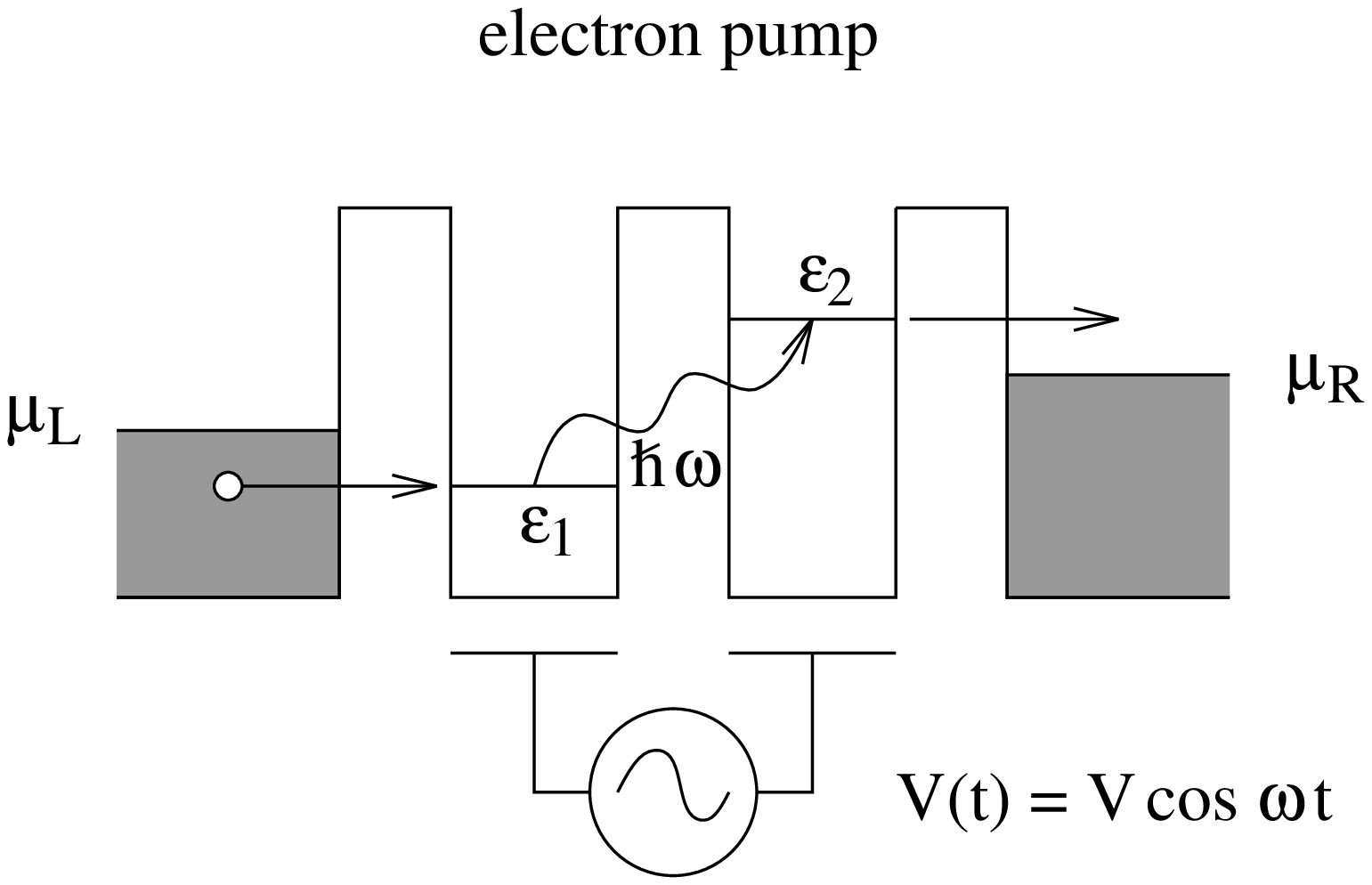}}
   \hss}
}
\caption{Schematic diagram of the double-quantum-dot electron pump.}
\label{fig1}
\end{figure}

\begin{figure}
\vbox to 10cm {\vss\hbox to 17cm
  {\hss\
    {\includegraphics{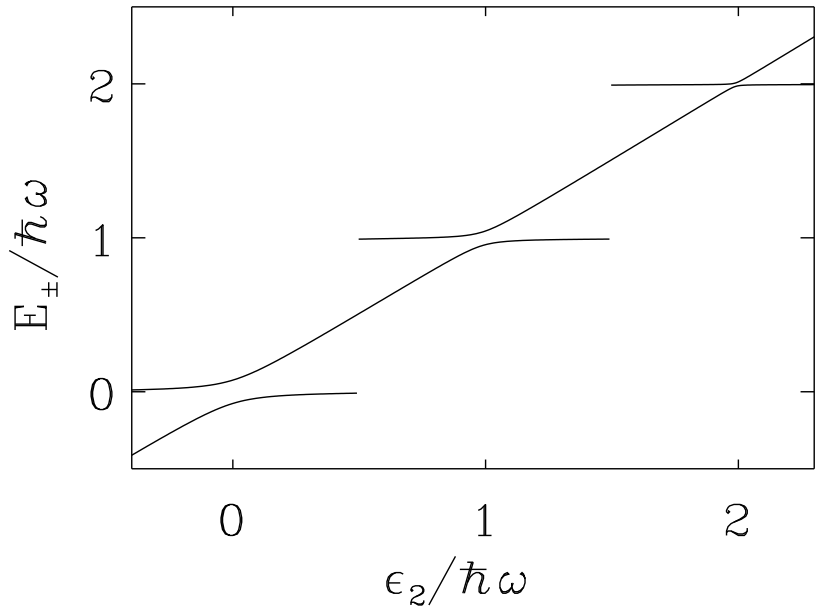}}
   \hss}
}
\caption{ Exact quasienergies of two coupled quantum dots
vs.\ detuning $\epsilon_2$.
Here $\epsilon_1 = V  \cos \omega t$, with $V=\hbar\omega = 10 \mbox{w}$.
Note that the quasienergies are defined $\mbox{mod} (\hbar \omega)$.
The electronic states on the dots hybridize and split by
$\simeq 2 \mbox{w} \mbox{J}_N (V/\hbar\omega)$, becoming delocalized,
when $\epsilon_2$ crosses the $N$th sideband of $\epsilon_1$.  }
\label{fig2}
\end{figure}

\begin{figure}
\vbox to 8cm {\vss\hbox to 17cm
  {\hss\
    {\includegraphics{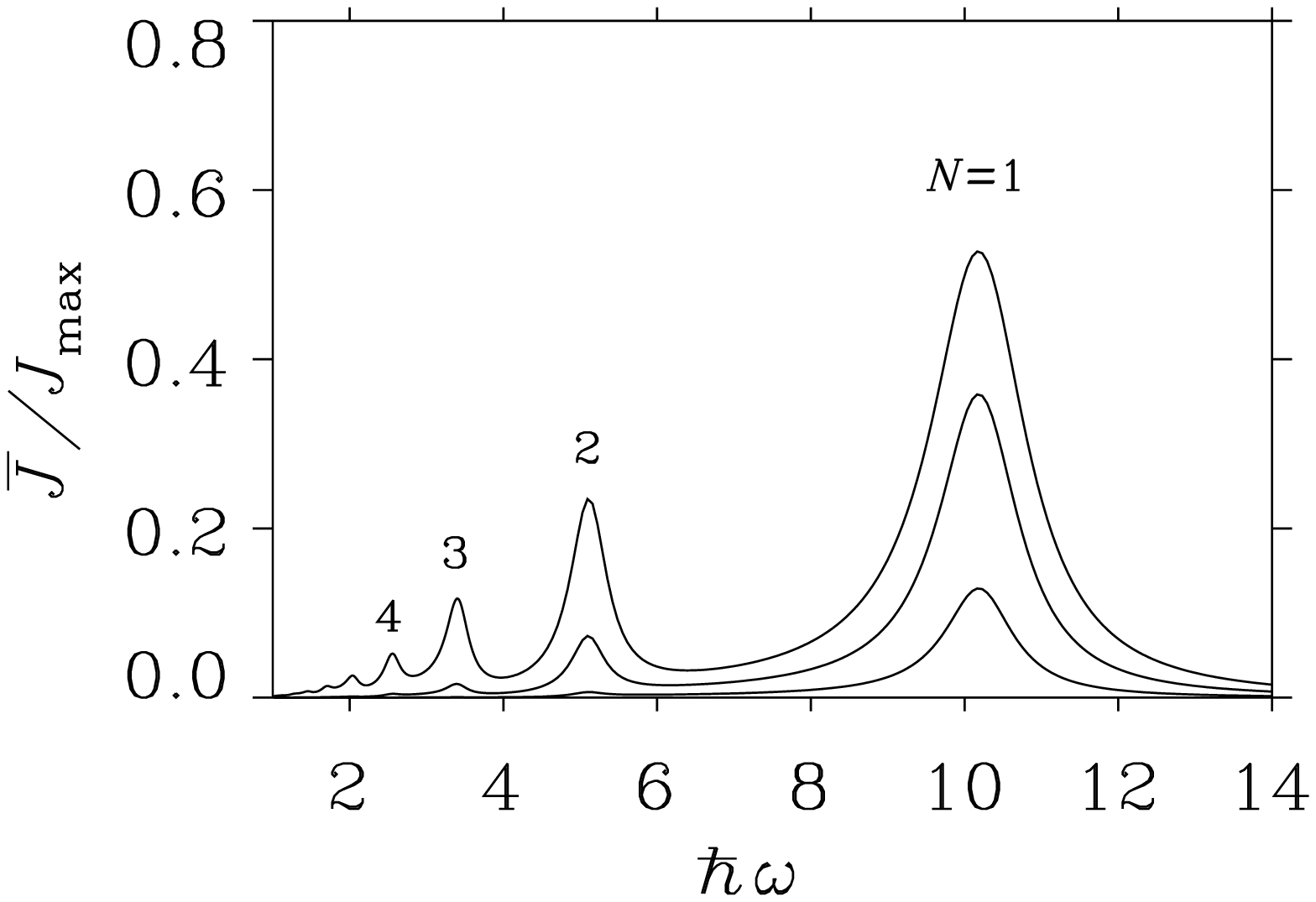}}
   \hss}
}
\vbox to 4cm {\vss\hbox to 17cm
  {\hss\
    {\includegraphics{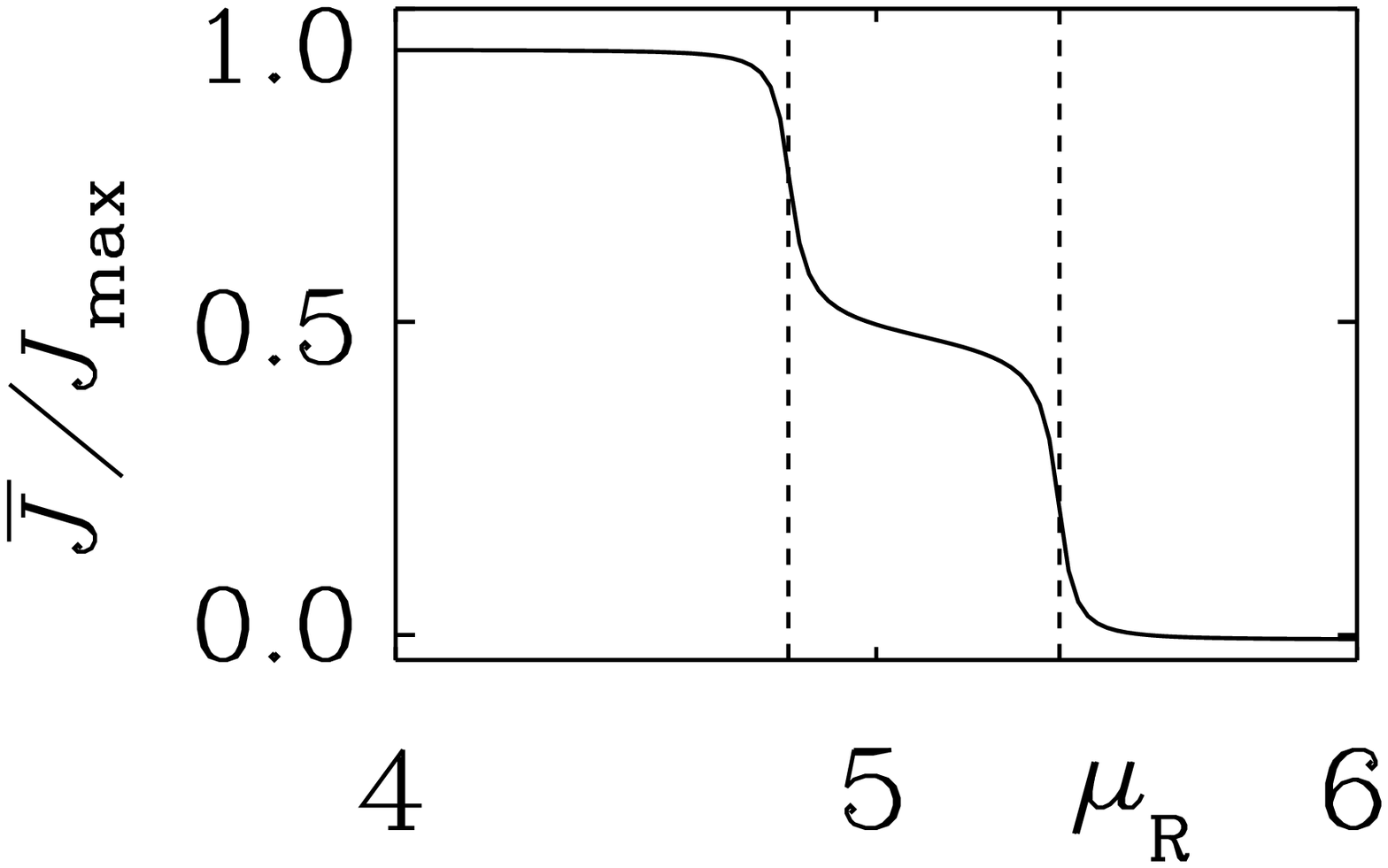}}
   \hss}
}
\caption{Time-averaged current $\bar{J}$
(in units of $J_{\mbox{max}}=e\Gamma/2\hbar$)
through a double quantum dot with
$\epsilon_1 =-5$, $\epsilon_2=5$, $\Gamma=0.5$,
and ac amplitude $V=2,4,6$ (increasing $\bar{J}$).  Energies are given in
units of $\mbox{w}$, the tunneling matrix element between the dots.
With $\mu_L=\mu_R=0$,
the system functions as an electron pump due to coherent $N$-photon-assisted
tunneling, with resonances at
$\omega_N = (\Delta\epsilon ^2+4\mbox{w}^2)^{1/2}/N\hbar$.
Inset: Time-averaged current
at the one-photon resonance $\omega= \omega_1$ versus dc bias
$\mu_R$, with $\epsilon_1 =-5$, $\epsilon_2=5$, $V = 6$, and $\Gamma =
0.05$. The vertical dashed lines indicate the values of the quasienergies,
$E_{\pm}$.  The jumps, of width $\Gamma$, of $\bar{J}$ at $E_{\pm}$
allow one to resolve the Rabi splitting $|E_+-E_-|
\simeq 2\mbox{w} \mbox{J}_1(V/\hbar\omega) = 0.563$.
}
\label{fig3}
\end{figure}

\end{document}